\documentclass[sigplan,screen]{acmart}

\AtBeginDocument{%
  \providecommand\BibTeX{{%
    \normalfont B\kern-0.5em{\scshape i\kern-0.25em b}\kern-0.8em\TeX}}}

\usepackage{multirow}
\usepackage{caption}
\usepackage{subcaption}
\usepackage{url}
\setcopyright{none}
\renewcommand\footnotetextcopyrightpermission[1]{} 
\begin{document}
\title{Rescuing the End-user systems from Vulnerable Applications using Virtualization Techniques}

\author{Vinayak Trivedi, Tushar Gurjar, Sumaiya Shaikh, Saketh Maddamsetty, Debadatta Mishra}
\affiliation{%
  \institution{\mbox{\textit{vinayaktrivedi7@gmail.com, tushar.gurjar1999@gmail.com, sumaiya@cse.iitk.ac.in, smaakd@gmail.com, deba@cse.iitk.ac.in}}}
  \country{Computer Science and Engineering, IIT Kanpur, India}
}


	\begin{abstract}
In systems owned by normal end-users, many times security attacks are mounted 
by sneaking in malicious applications or exploiting existing software 
vulnerabilities through security non-conforming actions of users. 
Virtualization approaches can address this problem by providing a quarantine environment
for applications, malicious devices, and device drivers, which are mostly used as entry 
points for security attacks. 
However, the existing methods to provide quarantine environments using 
virtualization are not transparent to the user, both in terms of application 
interface transparency and file system transparency. 
Further, software configuration level solutions like remote desktops and remote 
application access mechanisms combined with shared file systems do not meet the user
transparency and security requirements.
%

%
We propose qOS, a VM-based solution combined with certain OS extensions
to meet the security requirements of end-point
systems owned by normal users, in a transparent and efficient
manner. 
We demonstrate the efficacy of qOS by empirically evaluating the 
prototype implementation in the Linux+KVM system in terms of efficiency, 
security, and user transparency.
\end{abstract}

\def\UrlFont{\itshape}
\maketitle
\pagestyle{plain}
\section{Introduction}
\label{sec:intro}
%
%
%

The proliferation of computing end-points like laptops, smart-phones, and
other hand-held devices with internet connectivity at the fingertips of
gullible users pose security challenges of different kinds.
%
%
One such challenge is securing devices owned by end-users with little or no understanding of the
security intricacies of a complex and layered computing system. 
We refer to these device owners as {\em unwary users} in this document.
In systems owned by unwary users, security attacks are often mounted by maliciously sneaking in,
using security non-conforming actions of the unwatchful users.
For example, users may download malicious files from email or web applications and, despite system
generated warnings, allow the virus to take over the system by granting administrative privileges~\cite{istr}.
%
In this context, OS support for different user privileges~\cite{ostep} fails to protect the system as normal users being the owner of their systems also have the system administrator privileges.
%
Restricting the device owner privileges is not very user friendly and may raise usability concerns
in general. 
In this paper, we propose OS enhancements using virtualization technology to 
improve the security of
normal user systems in a {\em transparent and user-friendly manner}.
 
Virtualization techniques~\cite{xen, kvm, vmware, hyperv} provide support for strong isolation
across applications, used to intensify system security at different levels~\cite{secsurvey}.
Efficient sandbox techniques~\cite{minibox, mbox, proxos} to protect applications and the OS
from one another, isolation of potentially malicious devices and drivers~\cite{osdi2004, secvisor, copilot},
detection and mitigation using virtual machine introspection (VMI)~\cite{sigcse18, hindwai, arc2009, hprove}
are some example applications of the virtualization technique in the security context.
Further, hardware enhancements like Intel SGX~\cite{sgx} and Intel CAT~\cite{cat} can be used with
virtualization solutions to strengthen security~\cite{sgxvirt, scone, ryoan}. 
In the context of normal user end-point devices, hosted virtualization solutions like 
KVM~\cite{kvm} and HyperV~\cite{hyperv} can be leveraged to address security concerns by executing
applications or system components that are potential entry point of malware in isolated VMs.
For example, in Microsoft Windows10, the end-user is provided with a VM to run vulnerable applications in a sandbox using underlying Hyper-V hypervisor~\cite{win10}. 
%
%
%
While an ideal solution would entail the same {\em security} guarantees as that of VM-based solutions
along with the {\em efficiency} and {\em user experience} of a native system, achieving these three goals 
in their entirety and at the same time is non-trivial because of the following reasons. 

{\em First}, for any VM-based solution, it is difficult to remain {\em behind the scenes} and
still provide a cohesive view of the whole system.
%
%
%
Virtualization based approaches to provide secure computing platforms to unwary users
present two separate views of the file system and display interfaces (e.g., ~\cite{win10}).
For example, if a browser application is executed within a VM (for enhanced security) 
and a file is downloaded, accessing the file through the file system interface provided 
by the VM requires accessing the file system within the VM.
Similarly, to access the user interface of the applications (specifically GUI
applications), the user is required to access the VM console which may not be
very convenient.
{\em Second}, any security solution for unwary users should be simple to use and 
not involve additional actions from the user.
%
%
With existing VM-based solutions, the end-user is burdened and trusted 
to categorize applications and decide if a given application is required to be 
executed within the VM as this requires non-trivial expertise. 
For instance, a PDF viewer application may be treated as secure for all valid PDF files 
but becomes a potential security threat when opening a file downloaded through 
a browser or email client.
%
%
Further, if intelligent software agents on the machine take over the responsibility
of deciding the access to applications and files, the user should still be able to 
execute the applications (with some warning messages), instead of being completely 
restricted from using the application.
{\em Third}, the quarantine system should be efficient in terms of CPU and memory usage. 
While VM-based solutions are expected to result in certain overheads, the overheads should be as minimal as possible. 

Considering the challenges involving multiple dimensions, a practical solution 
should strive to reach a middle ground balancing the security, transparency, and 
performance aspects of the end-user systems.
Towards this objective, we present the
shortcomings of native systems and container-based solutions and, 
present two straw man (but unexplored in this context) 
VM-based solutions created combining techniques
like shared file systems and remote display access mechanisms (\S\ref{sec:background}).
%
Further, we motivate the need for a quarantine framework with fine-grained control over resource
exposure and enhanced security monitoring capabilities.  

We propose qOS, a set of OS extensions along with a VM sandbox
to enable a holistic quarantine environment which provides the user experience of a native system in a resource-efficient manner.
%
%
In qOS, a hidden VM (referred to as the qVM) is configured as the 
quarantine environment to take over the execution 
from the base machine when required (\S\ref{sec:design}).
Further, for end-user transparent file and display operations, 
the host OS provides a {\em controlled and monitored} channel
for applications executing in the qVM. 
For example, when an email client is executed from the base machine, 
the real execution happens within the
VM (without the user knowing it) and the user interacts with the application just like she would on a normal system.
Moreover, when the user downloads a file (e.g., text file) through 
an email client and tries to open it using
a text editor, the text editor also executes within the quarantine. 
Compared to simple VM sandboxes, qOS weakens the security in the quest 
for native-like user experience, but it offers
better security compared to other non-transparent VM-based
approaches and enables monitoring and filtering knobs in the framework (\S\ref{sec:security}). 
%
%
%
%

We evaluated our system in the Linux OS for different end-user applications like mail client ({\tt thunderbird}),
browser ({\tt dillo}), PDF viewer ({\tt evince}), text editor ({\tt gedit}), text-based browser ({\tt elinks}), 
HTTP client ({\tt wget}), secure text-based clients like {\tt ssh, scp} and {\tt sftp} (\S\ref{sec:expteval}).
The experimental analysis shows that qOS performs better than NFS-based straw man solutions 
in terms of CPU and memory efficiency. 
Further, while achieving the objective of being behind the scenes,
qOS improves display performance over remote desktop-based 
solutions by more than 15x. 
%
%
In short, qOS strikes a balance between the three design aspects as
compared to the straw man solutions; it offers better security guarantees and complete 
invisibility with lower resource overheads.

In this paper, we make the following contributions:

\begin{itemize}
\item Provide operating system enhancements to build a holistic quarantine environment using a hosted hypervisor to meet the security requirement of systems used by normal users who have very little expertise in computer system security. 
\item Design and implement a working prototype solution on Ubuntu Linux distribution with KVM hypervisor, which supports different GUI and command-line applications.
\item Empirically demonstrate the efficiency of qOS compared to partial alternate solutions.
\end{itemize}

\section{Motivation}
\label{sec:background}
According to a security study in 2018~\cite{istr}, a significant
number of attacks are performed exploiting the unmindful actions
of end-users, especially through internet activity.
%
To build secure systems for end-users, a threat model should be defined before 
exploring different solutions.

\noindent
{\bf Threat model:} For unwary user-owned systems, we assume at the time of installation, 
there are no malicious files in the system and, the vulnerability of installed applications
and the OS can be exploited by different types of malware
%
by entering the system through internet activity.
We also assume that the user is not a security expert and can be tricked
to gain administrator privileges from these malware.
With a typical VM sandbox, we assume that the hypervisor is the trusted
computing base (TCB) and can not be compromised through standard guestOS and 
hypervisor interactions.
%

\subsection{Why VM based quarantine?}
Native execution may allow malware to exploit vulnerabilities in the code through attacks like 
buffer-overflow~\cite{stacksmashing}, integer-overflow~\cite{intover}, return-oriented programming~\cite{rop,rop1,jop}
due to the large attack surface. 
Even though modern operating systems employ defense mechanisms like Address Space Layout Randomization(ASLR)~\cite{aslr}, 
stack guard~\cite{cowan1998stackguard} and Data Execution Prevention (DEP)~\cite{dep}, 
an unsuspecting user could inadvertently turn off these protections and allow malicious code to run on her system. 
For example, while installing any software on a device owned by an unwary user (with admin. privileges), 
the user may unknowingly agree to disable the ASLR feature compromising the security.
%
Moreover, malicious devices and device drivers can be potential entry points for attackers.
IOMMU based defense mechanisms to stop malicious devices~\cite{iommu} can not provide defense 
against buggy device driver code when 
the device driver runs with the OS privileges.
%

%

%

Containerization is a popular, lightweight solution employed for isolation and confinement~\cite{containers}. 
While containers provide namespaces for many subsystems including the file system, container-based solutions still depend on process level memory isolation techniques. 
This basic limitation of containers allows applications to break out of the container and gain root access on the host~\cite{docker1, docker2}. 
Virtualization provides a stronger form of memory isolation by enabling and managing separate address spaces
between the VMs and the hypervisor using techniques like virtualized memory management unit (MMU)~\cite{mishra2018survey, wang2011selective}.
Another advantage of virtualization over containerization is the flexibility at the hypervisor layer to enforce security policies, and employ detection and mitigation of security attacks~\cite{malwaresurvey, secvisor, rhee2009defeating, deterland}. 

\subsection{Possible VM-based solutions}
%
%

%



%
%
\begin{figure}[t]
 \centering
  \includegraphics[width=0.5\textwidth]{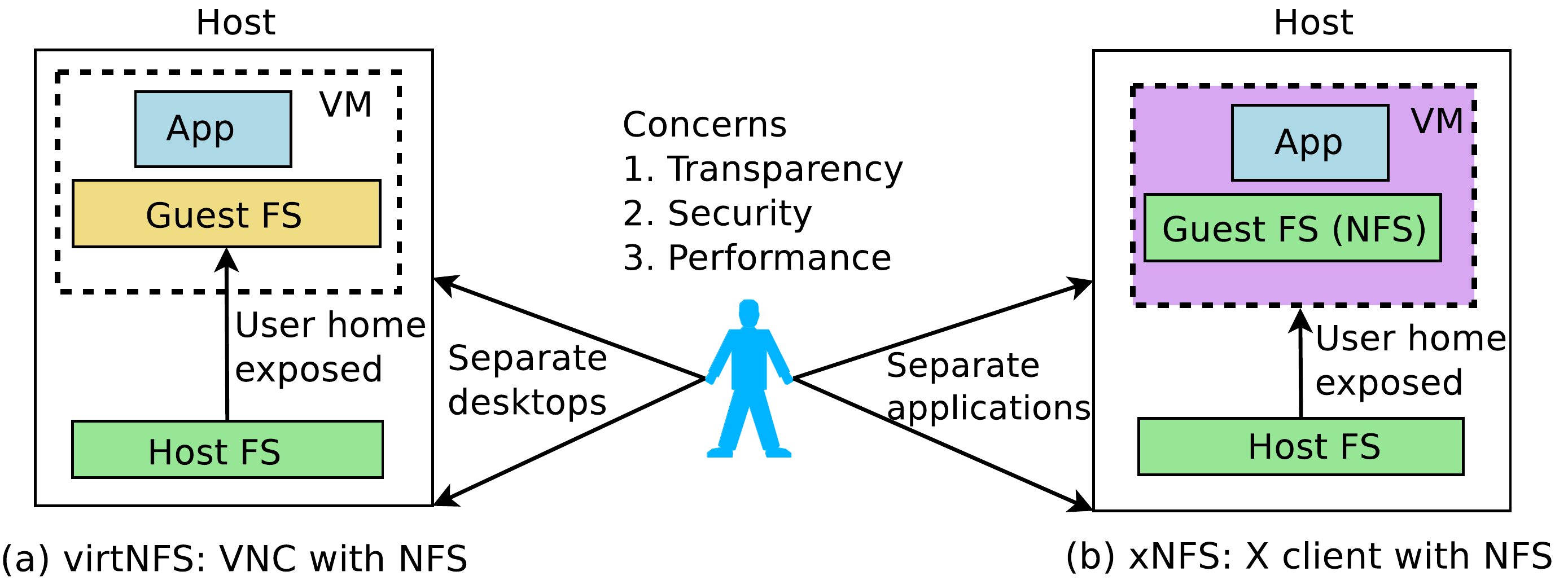}
  \caption{(a) Remote Desktop viewing system (b) Remote application interface with a shared file system.}
  \label{fig:approachcomparison}
\end{figure}

A simple VM as a quarantine environment like Windows sandbox or a VM
on a hosted hypervisor like KVM, apart from being overt, depends on the 
end-users for security enforcement.
Moreover, the end-user has two file system views and separate
display interface for the sandbox VM.
%
%
%
%
One possible approach to present a single file system view to the end-user 
is to use a network file system (NFS) from within the VM.
For display, two existing techniques---{\em (i)} Virtual Network Computing (VNC) 
based VM console access, {\em (ii)} X11 forwarding based application GUI access 
from the host---can be used. 
%
Figure~\ref{fig:approachcomparison} shows the two straw man approaches combining
the NFS with the display access techniques.
For file system transparency, user home is mounted 
in the VM to allow the user to share the 
same file system view (at least for the user's home directory).
We refer the NFS mounted scheme combined with VNC display access as 
$vncNFS$ and NFS mounted scheme combined with X forwarding based 
display access as $xNFS$.
%
$vncNFS$ can be thought as an extension to Windows Sandbox feature~\cite{win10} where
transferring the executable and its associated files to the sandbox 
is not required because of the NFS-based shared FS view.
Another possible alternate is to use light-weight VMs~\cite{vee2020, lightvm, lupine} 
to host applications but, even in this case, similar support for unified FS and display view 
is required.
%
%


%

\subsection{Why qOS?}

\begin{table}[t]
\small
\centering
\begin{tabular}{c|c|c|c}
\hline
    {\bf Design} & {\bf Sandbox} & \multirow{2}{*}{\bf vncNFS} & \multirow{2}{*}{\bf xNFS} \\
    {\bf goals} &  {\bf using VMs} & & \\
\hline
\hline
    Transparency  & \multirow{2}{*} {None} & Partial   & Better than \\
    (being hidden)&      & (only FS) & vncNFS \\
\hline
    \multirow{4}{*} {Security}     &  \multirow{4}{*}{Hypervisor} & Increased attack, & \multirow{4}{*}{Similar to} \\
                 &  \multirow{4}{*} {as TCB}     & surface, no in-built      & \multirow{4}{*}{vncNFS}\\
                 &             & monitoring and & \\ 
                 &             & control support& \\ 
\hline
    Performance  & Well studied & Unknown & Unknown \\
\hline
\hline
\end{tabular}
\caption{Comparison of solutions using VM as a sandbox.}
\label{tab:motivation}
\end{table}

The VM-based solutions presented above are not ideal as they do 
not satisfy all the requirements (See Table~\ref{tab:motivation}).
%
%
We lay down the aspects of an ideal solution and discuss the compromises made by 
$vncNFS$ and $xNFS$ which leads to the design objectives of qOS 
to achieve a better tradeoff.
%

\noindent
{\bf Transparency:} Ideally, the end-users should not be able to feel any difference in the way they interact with the system. Specifically, the desirable solution should provide a single file system view and present an application interface which is same as the interface in native execution.
The problem of user transparency remains as the user has to access 
two different machine interfaces (with $vncNFS$) and the application is launched 
from within the VM using the VM-level configurations and libraries (for both $vncNFS$ and $xNFS$). 
In qOS, we achieve better transparency where applications are
(re)launched in quarantine mode (if required) to provide the user experience
like a native system. 

\noindent
{\bf Security:} The ideal solution should be {\em as secure as}
a case of executing the vulnerable components within a VM where
the hypervisor is the TCB.
Additionally, the security should not assume end-user expertise and should be
non-restrictive i.e., the end-users may execute any arbitrary application
without compromising the host system security.
%
%
NFS sharing across the VM and the host has to be carefully designed such that 
the exposure is limited. 
Even with a carefully designed NFS, exposed host file system sub-tree when accessed
by other applications executing on the host can break the security.
For example, consider a case when a malicious PDF file is downloaded using the browser 
executing within the VM and saved onto the NFS share.
When the same file is opened from the PDF viewer executing natively on the host, 
the security is compromised.
Moreover, as the applications are treated as secure or vulnerable statically (irrespective
of the inputs) and executed within the VM or on the host, it will
have consequences with respect to security and efficiency.
qOS addresses these problems by separately classifying applications 
and inputs to the applications, not allowing any user-privilege change from
applications executing from within the VM and, enforcing fine-grained access control. 
%

\noindent
{\bf Performance:} The ideal performance is tricky as it can not be the same as the 
native system; an approximation can be to achieve the same performance of executing 
the application in the VM with a local file system and without any overheads due to 
display activities. 
While the virtualization overheads, in general, are well known, the performance overheads 
of a system like $virtNFS$ and $xNFS$ in terms of additional 
resource usage should be compared against qOS.

\section{Design of qOS}
\label{sec:design}
\begin{figure}[tp]
	\centering
	\includegraphics[width=0.39\textwidth]{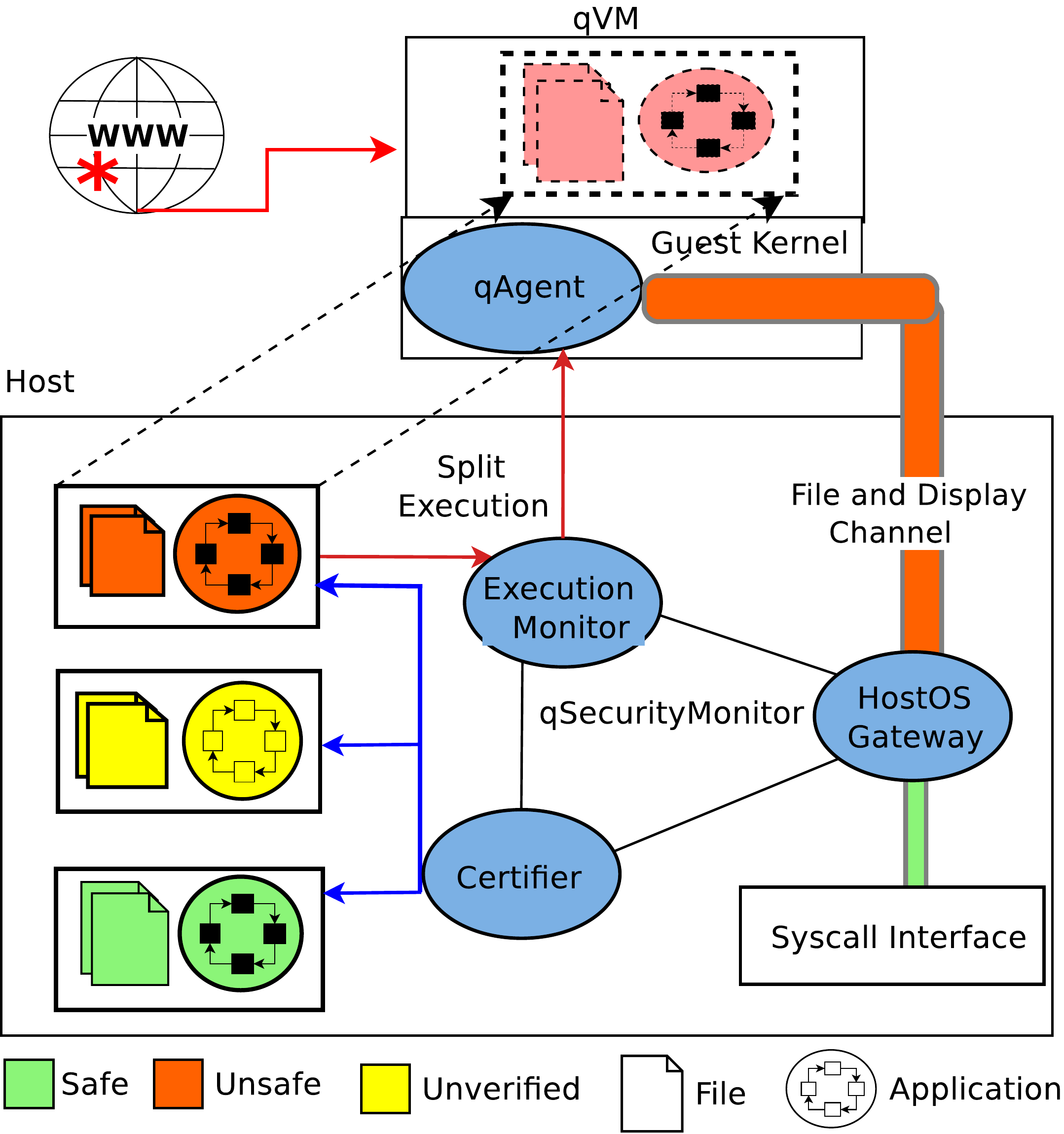}
	\caption{qOS architecture}
	\label{fig:new_design}
\end{figure}

Figure~\ref{fig:new_design} shows the high-level architecture of the proposed 
qOS system. 
The core design principle of qOS is to execute the untrusted and exposed applications
in a VM known as the quarantine VM (qVM).
To keep the host system secure, qOS employs a {\em safe access principle}
as explained below.
Applications and files are categorized into three categories---{\em (i)} safe (shown in green), {\em (ii)} unsafe (shown in orange) and,
{\em (iii)} unverified (shown in yellow).
Safe applications and files are allowed to be executed and accessed in the native mode on the host.
Unsafe applications are executed within the qVM and unsafe files are deleted
once they are detected.
Unverified files are created by unsafe applications during their execution and are not 
allowed to be accessed by any process in the host.
If the user must use an unverified file, the application is launched within the qVM.
Unsafe applications can be created by installing downloaded applications in the qVM and are
allowed to execute within the qVM.
qOS maintains and enforces the safe access rules at all times.
There are two major components in the qOS solution. 
On the host, qSecurityMonitor (qSM) is responsible for enforcing 
the security and execution policy.
In the VM, there is an agent (qAgent in Figure~\ref{fig:new_design}) which
communicates with the qSM to carry out application hosting and 
split-execution.

\noindent
{\bf Life cycle of qOS:} 
%
We assume that, when qOS is installed on the system, the host system is in a clean 
state i.e., there are no unsafe files in the system and applications are correctly 
categorized as safe and unsafe (no unverified files/applications in the system).
When the user tries to launch an unsafe application, the qSM dispatches the application
into the qVM through the qAgent.
The qAgent is responsible for the split execution by forwarding the file and display
related operations to the host through the qSM. 
The qSM monitors each request from the qAgent and updates the file and application
state (to unverified) if they are created/updated from the qAgent.
A periodic security analysis (by a security expert program and/or human) is performed on all the 
unverified entities to classify them as safe or unsafe.
At this point, if there are unsafe files, the security expert can analyze the genesis of the file by examining the qOS logs to decide further actions which may include reimaging the VM.
Moreover, if some applications are classified as safe, they may be installed on
the host to allow native execution afterwards.
Note that, the offline security analysis is a non-mandatory requirement (for efficiency) as
qOS design is flexible to execute applications and access files either from
the host or the qVM.



\subsection{qSecurityMonitor}
The qSecurityMonitor has the following subsystems:

\noindent
\textbf{Certifier}: The certifier is tasked with classifying entities 
as safe or unsafe for execution on the host system by employing two mechanisms. 
First, this unit is equipped with a configuration drafted by a security expert, 
based on which it takes the classification decisions. 
%
%
Second, it performs run time monitoring of operations performed by applications executing
within the VM to update the status of applications and files. 
One of the major challenges in enforcing the safe access principle is to restrict access to
the unverified (yellow) files from the processes executing on the host.
%
%
The certifier subsystem extends the file meta-data to store additional information
regarding the file classification state to address the above challenges.

%
%
%

\noindent
\textbf{Execution Monitor:} Execution of the unsafe entities
 is performed by splitting the processing across the qVM and the host. 
 The execution monitoring unit handles this bifurcation by co-ordinating with the qAgent.
 Unsafe applications can be launched in two different modes---{\em (i)} local mode (qOSL) and 
{\em (ii)} remote mode (qOSR).
 In local mode, the application binary and all libraries will be loaded from the qVM itself,
 while in the remote mode, the binary and libraries are loaded from the host system.
 The remote mode provides the flexibility of application launch and file access
 independent of their location.
 %
 %
 The execution monitoring is also responsible for handling process termination and host 
 state cleanup.

\noindent
\textbf{HostOS Gateway}: Split execution of applications within the VM requires
accessing the display and file system of the host to realize user transparency.
All the file and display operations are intercepted by the qAgent and forwarded to the
HostOS Gateway through the shared channel
(Figure~\ref{fig:new_design}). 
%
%
%
HostOS Gateway allows limited file/GUI APIs on the host side 
and closely monitors the robustness of each communication
by checking the parameters of each invocation.
Any non-confirming message from an application results in termination of the 
application along with a notification to the end-user.
At this point, the user may continue using the qVM or can perform a cleanup
and fresh installation.
%

\subsection{qAgent}
The qAgent subsystem is hosted in the qVM and provides two crucial functionalities to
realize split execution.
First, when the Execution Monitor requests to launch an application from within the VM,
the qAgent launches the application by loading it locally (in local mode) or by loading from
the host using the communication channel.
Second, the qAgent intercepts all file and display related operations originating from the
applications and forwards them to the HostOS Gateway through the communication channel.
In remote mode execution, qAgent has an additional responsibility of handling the 
dynamic library loading from the host.
%




\section{Implementation}
\label{sec:implementation} 

We have implemented a prototype of qOS in the Linux + KVM system where 
qVM is a Linux VM.
Both qAgent and qSM are implemented as kernel modules in the guest OS and host OS,
respectively.
Minor modification to the guest OS kernel is done to intercept the system calls.

\begin{figure}[!tp]
 \centering
  \includegraphics[width=0.47\textwidth]{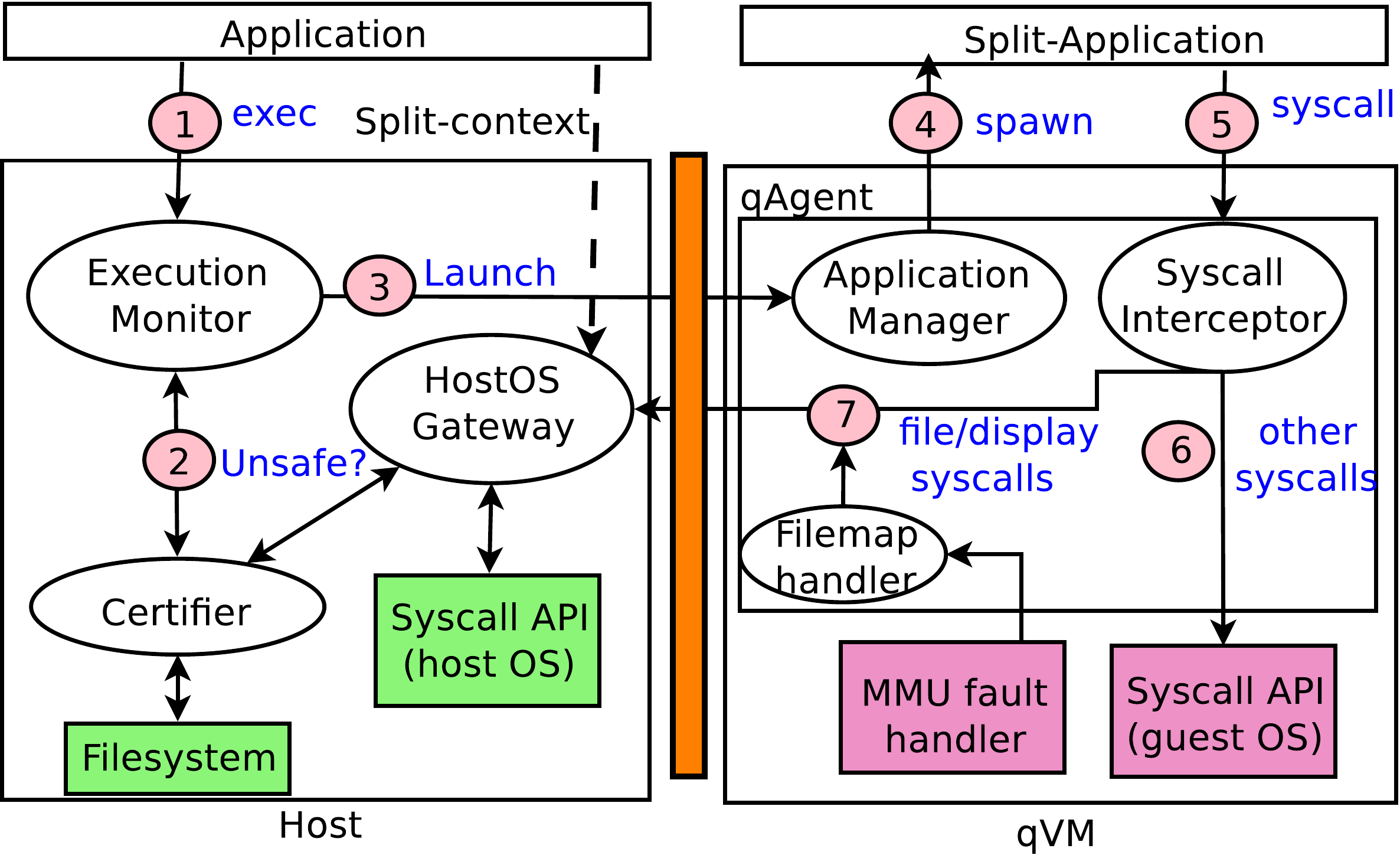}
	\caption{Implementation of split execution using qVM. Communication channel (shown in orange) is used
	         to pass messages between the qVM and the host. HostOS Gateway logic is
		 executed in the application process context.}
  \label{fig:implementation}
\end{figure}

Figure~\ref{fig:implementation} shows the working of different qOS components
when an unsafe application is executed from the host.
The Execution Monitor intercepts all {\tt exec} calls to find out (from the Certifier) 
if the application is unsafe and should be launched in the qVM.
Next, the Execution Monitor communicates (using the communication channel) with the application
manager (shown as ApplicationManager in Figure~\ref{fig:implementation})
to launch the application in the qVM by providing information like executable path, arguments and 
mode of execution (local or remote).
The process context on the host OS (shown as the Split-context) 
executes the gateway code where it waits to receive any
message from the qAgent.
When the application process executes in the qVM (shown as the Split-application), 
the system call interceptor (shown as Syscall 
Interceptor in Figure~\ref{fig:implementation}) forwards the file and display related 
system calls to the host OS gateway.
Similarly, if a file is mapped to the process virtual address space (using {\tt mmap}) and
a page fault occurs, the Filemap handler serves the request by communicating with the 
host OS gateway.
The syscall interceptor maintains the correlation with the host file descriptors
to handle subsequent operations.
When any file is opened in write mode, the HostOS Gateway notifies the Certifier
to update the file category information (Refer \S\ref{sec:design}).
Once the split-application within the VM exits, the split-context on the host is cleaned up.

\subsection{File and display channel} 
Whenever the split-application makes an open system call, 
the system call interceptor decides whether to open the file in host or qVM. 
Devices, sockets, pipes and libraries (in local mode) are opened in qVM itself. 
For all other regular files, the system call interceptor sends a request to 
the split-context waiting in the host OS gateway. 
The host OS gateway opens the file in the host and returns the file handle 
to the system call interceptor.
The system call interceptor maintains a per-process mapping of the 
file handles to serve the system calls like {\tt read}, {\tt write} etc. 
%
%
Operation on the standard file descriptors ({\tt stdin}, {\tt stdout} and
{\tt stderr}) are always forwarded to the host.
Similarly, for GUI display operations, all the X client socket operations are forwarded to the 
host.
%
%
In the current implementation of qOS, there is a single process instance (split-context) 
in the host side for each application irrespective of the number of processes/threads 
in the qVM.
On the qVM side, the system call interceptor is executed in the context of the process/thread
that invokes the system call.
Therefore, when a process/thread is created/terminated in the split-application, 
the file reference information is updated in the split-context on the host.
The primary reason for using an N:1 design is security i.e., 
to avoid DoS attack on the host (See \S\ref{sec:discussion}). We avoid blocking I/O calls and use adaptive I/O event handling to perform I/O multiplexing.

\vspace{0.1cm}
\noindent
{\bf File mmap:} Mapping a file into the process address space using {\tt mmap} system 
call requires special handling if the file belongs to the host system. 
One of the strategies could be to populate the address space by reading the whole file 
at the time of mapping which may result in slow {\tt mmap} performance.
qOS adapts an on-demand strategy by registering a fault handler for the 
filemap region and serves the page faults by reading the file content 
from the host.
We have observed that for all applications most {\tt mmap} calls 
are used to map the dynamic libraries into the address space and only 
a part of the library is actually used during the execution. 
It is possible to design more sophisticated schemes (e.g., populating
pages in the vicinity of fault address) and left as a future direction.
 
\vspace{0.1cm}
\noindent
{\bf I/O multiplexing:} Event driven I/O using {\tt select} and {\tt poll} require
to be split carefully.
When the file descriptor set (FDSET) elements are all 
local fds (like pipe, network connections etc.) or all host fds, then the handling is 
trivial.
However, when the FDSET is a mixture, we have to wait for the I/O event in both the 
host and the qVM.
%
%
%
We have designed an adaptive strategy to invoke the guest OS and host system calls 
in turns with an appropriate time slice determined dynamically based on the 
past behavior and application profile.
While we have implemented an adaptive strategy for the {\tt select} system call, 
for {\tt poll} system call we juggle between the guest OS and the host 
with a fixed time slice of 10ms.

\subsection{Communication channel}
The communication channel is implemented using the shared memory
region between the qSM and the guest OS enabled through
\textit{ivshmem}~\cite{macdonell2011low}.
The shared memory region is divided into two types of 
message slots---{\em normal slots} and {\em big slots}.
Each normal slot is implemented as a circular queue for message passing.
Each process/thread of a split-application is allocated two normal slots
for communication in each direction.
This allows lock-less operation on the slots for multiple multithreaded
split-applications.
The big slots are used to pass large messages (like binaries, file data etc.)
and they need to be reserved before use.
The slot allocation is performed by the execution monitor on the host side 
at the time of application launch or process/thread creation.
The split-context on the host can wait on multiple slots to handle messages
from any of the threads/processes in the qVM.

\subsection{qSM: safe access principle}
%
If the application running in the qVM creates/updates any file on the host system, 
the qSM does not allow the file to be opened or executed on the host system. 
qSM achieves this functionality by leveraging the extended file attribute feature 
of the Ext4 file system. 
Whenever a file is written from the split-application, 
qSM sets the extended attribute to reflect that it is unverified. 
We have modified the VFS layer to stop any process from opening the file in the host.
A subtle (but rare) case arises when a file already in use by a process on the host is modified
by a process executing within the qVM. To tackle this case, we verify 
the file object references in the in-memory inodes to ensure that the file
requested to be opened for writing from the qVM is not in use by any 
other process on the host.

\subsection{Subtle issues and optimizations}
\label{sec:subtle}
We encountered and addressed many corner cases during the implementation of 
qOS. Some of them are highlighted here.
First, killing a split-application from the host using signals (e.g., CTRL+C) 
requires the kernel signal handler invocation at both the host and the guest OS.
Second, many split-applications use {\tt clone} system calls with 
arguments which does not match any related POSIX calls like {\tt fork},
{\tt vfork} or {\tt pthread\_create}. To tackle this, qOS relies on
{\tt CLONE\_FILES} to update the file usage count on the host side.
Third, some applications use UNIX socket control messages~\cite{cmsg} 
to setup file descriptors in other processes which need to be
tracked by the qAgent. 
Fourth, the {\tt CLOSE\_ON\_EXEC} flag is set/unset using different system 
calls ({\tt open}, {\tt socket}, {\tt fcntl} etc.) which should be 
tracked to avoid file descriptor leakage on the host.
Fifth, it may happen that an application (safe) running on the host, 
due to some user input or otherwise, tries to open an unverified file. 
In such a case, the file access is prevented and the user is informed of the unsafe access.
There is a provision in qOS to temporarily mark such an application unsafe 
and re-execute the application to launch it using qVM (if required using the
remote mode). 

One of the primary optimizations is to minimize the extensive communication between system call interceptor and the host OS
gateway.
%
To accomplish this, we employ techniques like write-combining and read-batching. 
In write-combining, qAgent uses a big slot to store many write requests on the same file and sends a single write request. In read-batching, 
the system call interceptor in qAgent reads bigger chunks
to serve subsequent read requests from the split-application~\cite{10.1145/191080.191133}. 
Note that, these optimizations require careful handling of consistency issues caused due to
simultaneous writes.
In simultaneous write scenarios, the optimization is turned off and normal operation is
carried out.

\section{Security Analysis}
\label{sec:security}
\begin{table*}
\small
\centering
\begin{tabular}{c|c|c|c|c}
	{\bf Threat} & {\bf Simple VM} & {\bf vncNFS} & {\bf xNFS} & {\bf qOS}	\\
\hline
\hline
	Virus entry due to     &  \multirow{2}{*}{Contained} & Vulnerable & Vulnerable & Contained \\
	unwary user actions    & & (malicious files via NFS) & (malicious files via NFS) & (qSM monitoring) \\
\hline
	Control flow hijack  & \multirow{2}{*}{Contained} &  Contained (assuming VNC & Contained (assuming X & Contained (assuming \\
	(e.g., buffer overflow) & & and NFS are bug free) & and NFS are bug free) & X is bug free) \\
\hline
	Denial of service & Host is  & DoS through & DoS through & Impacted, but has \\
	(on the host)     & not impacted & the NFS server  & the NFS server & knobs for detection\\
\hline
        Information leakage & \multirow{2}{*}{None} & None & Vulnerable & Vulnerable, extended \\
        using display (e.g., clipboard) & & (separate display) & (additional config. required) & monitoring needed\\
\hline
         Information leakage & \multirow{2}{*}{None} & Vulnerable & Vulnerable & Vulnerable \\
         using file system channel & & (NFS exposed files) & (NFS exposed files) & (Files allowed to the user) \\
\hline
\hline
\end{tabular}
\caption{Comparative security analysis of qOS vis-a-vis other VM-based solutions.}
\label{tab:security}
\vspace{-0.7cm}
\end{table*}

In this section, we present a comparative security analysis across four different
quarantine environments---{\em (i)} VM as a quarantine with local file system and
display, {\em (ii)} $vncNFS$ (explained in \S\ref{sec:background}), {\em (iii)} $xNFS$ (explained
in \S\ref{sec:background}), and {\em (iv)} qOS.
Table~\ref{tab:security} shows the overview of security threats and their implications
on different setups.
Note that, the threats shown in Table~\ref{tab:security} can impact the native systems
easily compared to all the other setups.
%

\noindent
{\bf Virus entry to host FS:} One of the primary design objective of qOS is to address the security issues caused due to the
actions of unwary users.
A malicious entity (e.g., malware, trojan etc.) can enter into the system because the
user can accidentally download it through a browser or email client.
qOS provides the same level of security as the simple VM setup as qOS prevents
any application from host accessing the malware.
%
The $vncNFS$ and $xNFS$ setups can be penetrated because once the file is downloaded and 
accessible natively, the host system becomes vulnerable.
A compromised application from the qVM can corrupt accessible files as neither qOS nor a
trusted VFS layer enforcing the user ACLs correctly can stop any legitimate write access.
To address this kind of attacks, intelligent monitoring techniques proposed 
in gVisor~\cite{gvisor}, MBOX~\cite{mbox} etc.
can be incorporated into the qOS framework.

\vspace{0.1cm}
\noindent
{\bf Control-flow hijack:} The attacker exploits the bugs or vulnerabilities
in applications by passing carefully crafted inputs to gain access to the system.
In the quarantine execution model, all the attacks are localized to the VM. However,
both $vncNFS$ and $xNFS$ assume the NFS server to be bug-free such that even if the 
NFS client in the VM is compromised, NFS server at the host remains unimpacted.
Further, both $xNFS$ and qOS use the X-server executing on the host to serve the
requests and assume that the X-server to be secure.
For qOS, the VFS layer is trusted to enforces the user
ACLs correctly which allows qOS to detect applications trying to
access files without having necessary permissions.

\vspace{0.1cm}
\noindent
{\bf Denial of service:} With xNFS and vncNFS, any malicious application in the 
VM can perform a lot of file operations to impact the 
applications executing on the host.
%
For example, a compromised process in the VM can create and delete 
files on the host in an infinite loop.
This is particularly a difficult attack to stop in general.
%
The qOS framework can identify the applications with abnormal
behavior (at the HostOS Gateway) to rate-limit 
such operations and in the worst case terminating the application
or re-imaging the qVM.
%

%

\vspace{0.1cm}
\noindent
{\bf Information leakage:} Due to complete separation of display and file systems, 
information leakage from the host is not possible in a simple VM setup.
$xNFS$ and qOS (in the current implementation) can leak information through the
display channel by allowing access to resources like a clipboard. 
qOS security monitoring can be extended to stop access to these resources 
as the split interface is suitable for such an extension.
While qOS does not allow any user privilege change, information leakage from 
accessible files is hard to stop in all the setups as any file 
on the host with access permission for the user can be accessed.

\noindent
{\bf Other attacks:} Attacks like micro-architectural
side-channel attacks are equally applicable for all the setups.
Compared to native execution, this attack is difficult to perform on the
virtualized system because of the noise. 
Moreover, hardware defense mechanisms
like cache allocation techniques and, hardware enclaves can harden the 
system further.

\noindent
{\bf Security of qOS channels:} qOS implements the communication channel
by sharing memory across the host and guest OS. 
Considering the worst-case scenario when the guest OS is compromised and 
the communication channel is used maliciously, qOS is capable of
filtering out messages which can impact the host system.
The messages sent through the communication channel follow a very strict 
format and any violation of the format can be detected at the host.
%
%
Any message not meeting the message structure is discarded and the application
is killed.
Note that, even if the communication slots for applications are isolated,
a compromised qVM OS can send legitimate messages on slots used by other
applications.
In such a case, the impacted applications will malfunction which acts as an 
indication to re-image the qVM.
%
%
%

%
While qOS provides a more security hardened solution compared to $xNFS$ and
$vncNFS$, it compromises certain aspects of security of a VM-only solution.
We believe that, qOS security can be improved by employing complimentary 
techniques like VMI to effectively monitor the guest OS and the 
communication channel.

\section{Experimental Evaluation}
\label{sec:expteval}

{\bf Experiment setup:} To evaluate the efficiency of qOS, we performed our experiment using 
Linux Ubuntu-14.04LTS desktop on Intel i7 machine (eight logical CPUs with hyper-threading) 
with 16GB RAM.
A VM (using KVM hypervisor) with two logical CPUs (pinned to physical CPUs) with 2GB RAM is used as the quarantine
environment in our experiments (referred to as qVM).
Our prototype implementation of the proposed quarantine environment using qOS extensions 
are implemented in the Linux kernel version 4.20.0 (both in the VM and the host).

\begin{table}[t]
\small
\centering
\begin{tabular}{c|c|c}
\hline
{\bf Setup} & {\bf Details} & {\bf GUI access method}\\
\hline
\hline
    {$NoSec$} & Host system, no isolation & Local \\
    {$vncNFS$} & VM with NFS mount & VNC viewer \\
    {$xNFS$}  & VM with NFS mount & SSH + X\\
    {$qOSL$}  & qVM local mode & Local \\
    {$qOSR$}  & qVM remote mode & Local \\
\hline
\hline
\end{tabular}
\caption{Experiment configurations.}
\label{tab:setup}
\end{table}

For comparative analysis, we have used five different setups, as shown in the Table~\ref{tab:setup}. 
The $NoSec$ system is the host system without any virtualization-backed quarantine.
$vncNFS$ mode requires the user to access the VM desktop through VNC viewer while
$xNFS$ mode allows opening applications directly using X-forwarding (e.g., ssh -X).
$vncNFS$ and $xNFS$ modes use a VM where we mount the home directory of the host machine 
to a VM directory and use the mount location (whenever possible) to approximate transparent 
file system view.
Note that, both these modes still use the VM file system for accessing configuration files,
application caches, etc. from the VM user's home directory and system directories.
The two qOS modes ($qOSL$ and $qOSR$) are identical in all aspects but differ in the 
manner they load application binaries and libraries, where the former loads the binary and 
libraries from the qVM while the latter loads it from the host.
The network interface is assigned to the VMs through VMM bypass~\cite{vmmbypass, sriov} for 
all configurations except $NoSec$. 
Additionally, $xNFS$ uses an internal network interface (exposed through VirtIO~\cite{virtio}) for X
communication between the VM and the host.

\begin{table*}[t]
\small
\centering
\begin{tabular}{c|c|c|c|c|c|c}
\hline
\hline
\multirow{2}{*}{\bf Workload} & \multirow{2}{*}{\bf Operation} & {\bf \# of qOS } & 
    {\bf \#of mmap } & {\bf \# of display I/O } & {\bf \# of File I/O } & {\bf I/O event} \\
    & & {\bf messages } & {\bf messages} & {\bf messages} & {\bf messages} & {\bf messages} \\
\hline
    \multirow{4}{*}{Email} & Search ($qOSL$) & 6146 & 3 & 2840 & 110 & 2999 \\ 
    & Search ($qOSR$) & 6674 & 495 & 2563 & 151 & 2913 \\ 
    \cline{2-7}
    & Send email ($qOSL$) & 25787 & 12 & 9513 & 3167 & 9306  \\ 
    & Send email ($qOSR$) & 27457 & 996 & 7908 & 5444 & 6801 \\ 
\hline
    \multirow{4}{*}{Web browser} & MultiSite ($qOSL$) & 8051 & 0 & 3819 & 1241 & 1905 \\ 
    & MultiSite ($qOSR$) & 60452 & 39707 & 3301 & 2731 & 1645 \\ 
    \cline{2-7}
    & Download ($qOSL$) & 5971 & 0 & 3454 & 97 & 1933 \\ 
    & Download ($qOSR$) & 10751 & 2324 & 3589 & 407 & 1992 \\ 
\hline
    \multirow{2}{*}{PDF viewer} & Search ($qOSL$) & 27210 & 0 & 5326 & 16686 & 5198 \\ 
    & Search ($qOSR$) & 29222 & 119 & 5679 & 17057 & 6477 \\ 
\hline
    \multirow{2}{*}{Text editor} & NewDoc ($qOSL$) & 33102 & 0 & 16442 & 16 & 16403 \\ 
    & Search ($qOSR$) & 34555 & 383 & 17022 & 279 & 16113 \\ 
\hline
\hline
\end{tabular}
\caption{Breakdown of host operations performed for different application usage scenarios for qOS local and remote modes.}
\label{tab:insights}
\vspace{-0.5cm}
\end{table*}
\subsection{GUI application overhead}
{\bf Experiment methodology:} We use popular end-user applications {\tt thunderbird}~\cite{thunderbird}
email client, {\tt dillo}~\cite{dillo} low footprint web browser, GNU editor ({\tt gedit}) and PDF 
viewer for our experiments.
One of the challenges in the empirical analysis of GUI application performance is the {\em human factor}
during the experiments.
We use an automated input generation mechanism provided by the Linux kernel community~\cite{uinput}
to trigger application actions without any human intervention. 
For example, to send an email using the {\tt thunderbird} email client, we generate the key shortcuts and
mouse movements required to send an email using an input generation program while the {\em thunderbird} application 
is the active window in the system.
We repeated the same key generation pattern after starting the application using different quarantine setups
and captured system statistics to analyze their performance comparatively.
All GUI experiments were repeated three times, and the average readings are reported.

\subsubsection{Email client}

\begin{figure}[tp]
\centering
\includegraphics[width=0.85\columnwidth]{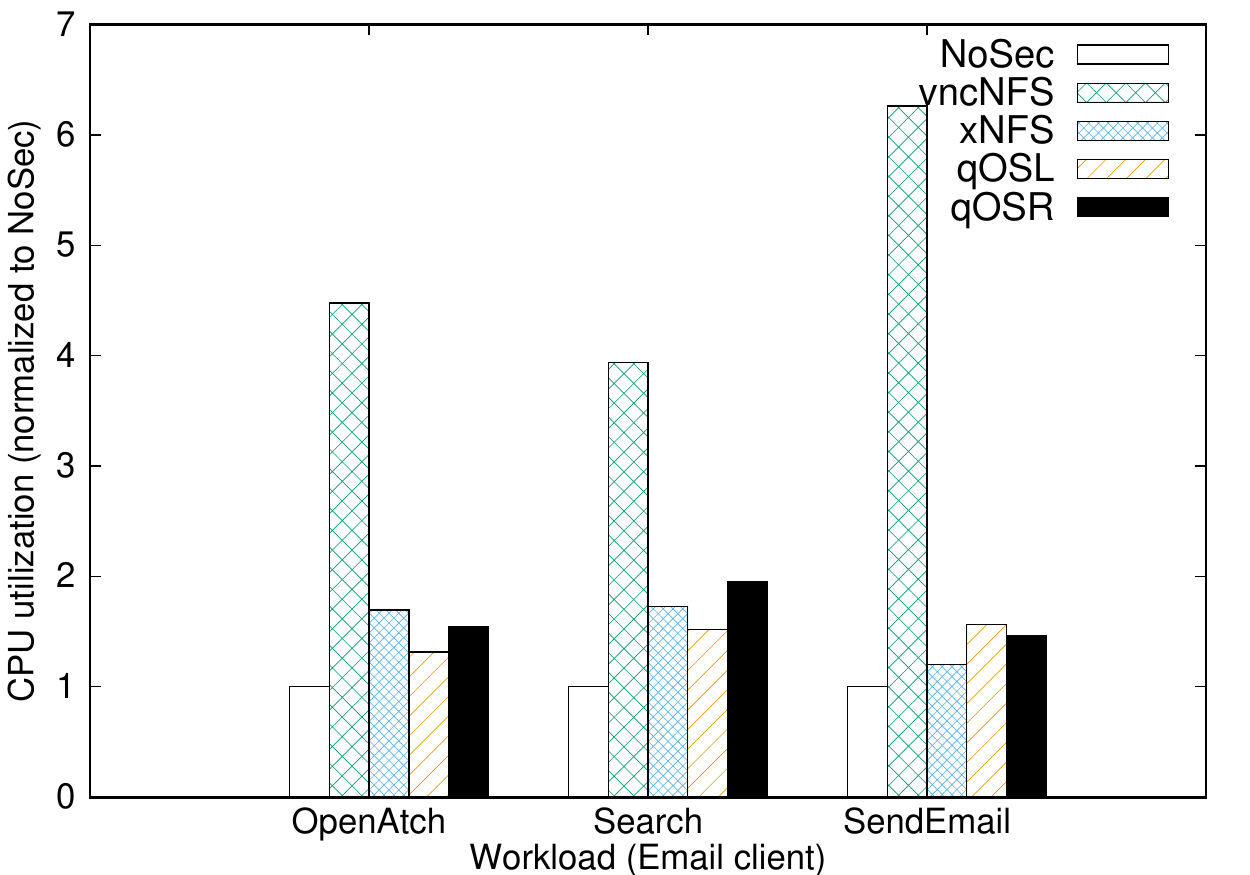}
    \caption{Comparative CPU overhead of different operations using the {\tt thunderbird} email client.}
\label{fig:thunder}
\end{figure}
For this experiment, we used {\em thunderbird} email client configured for an existing account in an email server hosted over LAN
containing a huge number of emails (more than 15000). 
We used three workload profiles for this experiment as shown in Figure~\ref{fig:thunder}---{\em(i)} open a PDF attachment
from an email (OpenAtch), {\em(ii)} search a string in the emails (Search), and {\em (iii)} send an email to the 
self account. All these operations are performed through the automated input mechanism explained above.
For $vncNFS$ and $xNFS$ modes, we have used the same email account.

As shown in Figure~\ref{fig:thunder}, $vncNFS$ results in maximum overhead compared to all other modes---4.5x, 4x and more than
6x compared to $NoSec$ mode for open attachment, search and send email, respectively.
In terms of CPU utilization, $qOSL$ performs better than $xNFS$ in open attachment and search workload scenarios by a factor of 
1.29x and 1.14x, respectively.
CPU utilization for sending an email is better in case of $xNFS$ compared to both $qOSL$ and $qOSR$ by $\sim$ 1.25x.
This can be attributed to excessive I/O event messages ({\tt poll}) ($\sim$30\% of total messages) 
between the Syscall Interceptor (in qVM) and the HostOS Gateway (on host) during email editing. 
As we have discussed earlier, we have not implemented {\em adaptive select} logic for {\tt poll}
and the {\tt poll} performance can be improved by such a design.
One more possible factor can be the file I/O activity performed on the host while in case of
$xNFS$, it is performed on the local file system.
$qOSR$ results in higher CPU usage by a factor of 1.17x and compared to $qOSL$ for email attachment and search operations.
This is primarily due to the increased number of {\tt mmap} faults in case of $qOSR$ (Table~\ref{tab:insights}) that are 
handled by performing a fault handling over the communication channel.

\subsubsection{Web browser}
\begin{figure}[tp]
\centering
\includegraphics[width=0.85\columnwidth]{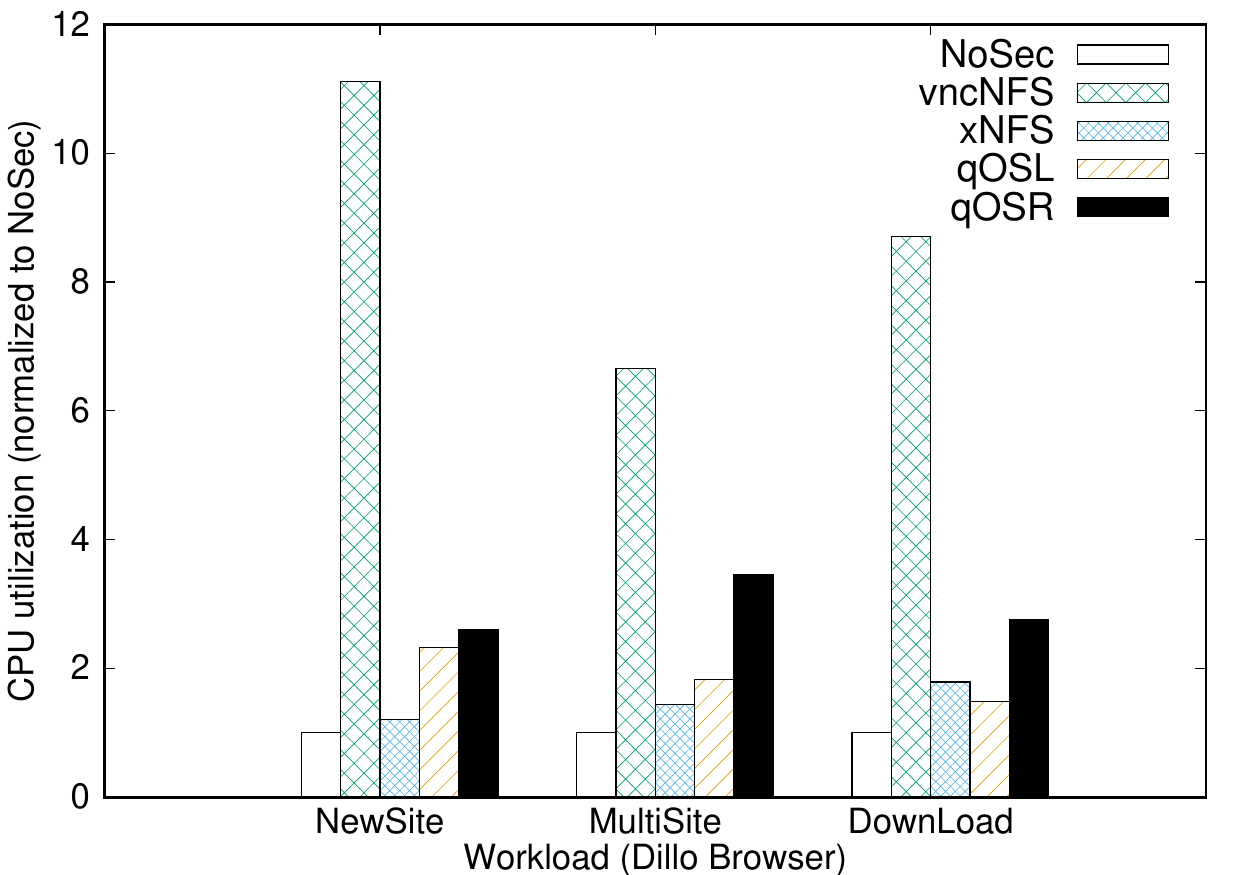}
    \caption{Comparative CPU overhead for web browser ({\tt dillo}) workload.}
\label{fig:dillo}
\end{figure}

For this experiment, we used {\tt dillo} web browser with three different workload profiles.
As shown in Figure~\ref{fig:dillo}, the workload profiles are: {\em(i)} open google search page (NewSite), {\em(ii)} open multiple 
web sites using tabs (MultiSite), and {\em (iii)} download a small PDF file of size 300KB (and store it on NFS mount for $xNFS$ and $vncNFS$).
For opening a simple web page like google search page, the absolute CPU utilization was very low (between 0.25\% to 1\%) 
for all workloads except $vncNFS$, where the CPU utilization was around 4\%.
$vncNFS$ results in more than 6x and 8x additional CPU overheads compared to the $NoSec$ system 
for MultiSite and Download scenarios, respectively.
$qOSL$ performs better than $xNFS$ for Download (by 1.2x) while results in additional CPU overheads
by a factor of 1.25x for Multisite scenario.
Comparatively, $qOSR$ results in higher overhead in case of Multisite workload 
than the Download workload.
This can be explained by observing the file I/O activity across the two workload scenarios where with MultiSite, the number of I/O operations are little less than 7 times compared to the Download workload (See Table~\ref{tab:insights}).

\subsubsection{PDF viewer and text editor}
\begin{figure}[tp]
\centering
\includegraphics[width=0.85\columnwidth]{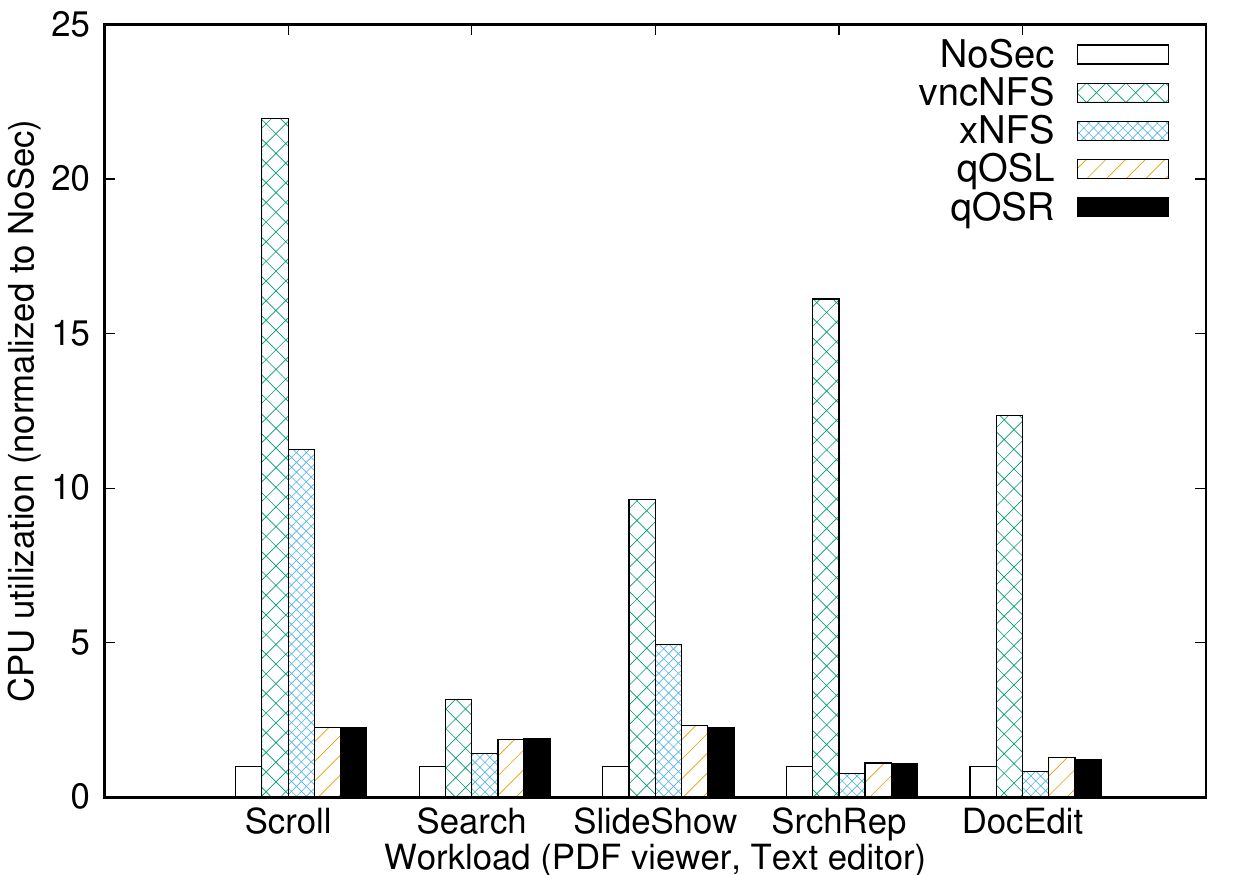}
    \caption{Comparative CPU overhead for PDF viewer ({\tt evince}) and Text editor ({\tt gedit}) workloads.}
\label{fig:evince}
\end{figure}

In Figure~\ref{fig:evince}, we present the performance comparison of two workloads with the following
application scenarios. For PDF viewer, we used a 7MB PDF document (on the NFS for $xNFS$ and $virtNFS$  modes) 
and performed three operations i.e., {\em(i)} scroll the PDF file using down arrow (referred as Scroll), {\em (ii)} search a frequently appearing key word
in the document (Search) and, {\em(iii)} perform slide show for 50 pages (SlideShow).
To evaluate text editor, we used two workload profiles---search and replace a common keyword from an opened document (referred to
as SrchRep) and edit a new document by entering 1000 alphabets and saving it (DocEdit).

As shown in Figure~\ref{fig:evince}, $xNFS$ performs marginally better than qOS for 
PDF Search while results in significant overheads (around 5x and 2x) compared to
the qOS modes for Scroll and Slideshow, respectively.
$vncNFS$ results in lower performance overheads in Search operation compared to other 
operations as the workload CPU usage is primarily due to the application code and 
file read operations (from the local page cache).
Even with the read buffering scheme, the qOS mode performance is still impacted because
of heavy file I/O (around 60\% operations on the host are file I/O operations as shown in Table~\ref{tab:insights}) 
can not match the local caching performance achieved because of NFS.

For Text editor workload ({\tt gedit}), CPU overhead for $vncNFS$ is significantly
higher than the other modes (up to a factor of 15). CPU overheads for all other modes
are comparable to the baseline $NoSec$ setup.
Interestingly, for editing a new document (DocEdit), the performance of xNFS mode is
better than $NoSec$ setup.
We observed the same behavior with repeated experiments and found that the Text editor
loads some profiles from the local system, and this may be different in the qVM and the
host.
The performance overhead of qOS modes with DocEdit (by 1.29x more CPU usage compared to $NoSec$)
can be explained by excessive event I/O requests (similar to {\tt thunderbird}) and display
related I/O requests, when added contribute more than 90\% of the total requests (Table~\ref{tab:insights})
sent across to the host system.

\noindent
\textbf{Summary:} 
The performance of vncNFS is not very promising while xNFS with its limitations results in low
performance overheads in all applications except PDF viewer. Between xNFS and qOS, there is no clear winner with respect to resource overheads, better qOS modes are consistent across applications and offer
other advantages as we show in the next subsection.

%

\subsection{Command line application performance}
We have used applications of two categories for the following experiments---batch mode network operations like
downloading big files, directories and websites, and, interactive operations on a remote shell and
text-based browsing.

\subsubsection{Batch operations}
\begin{figure}[tp]
\centering
\includegraphics[width=0.85\columnwidth]{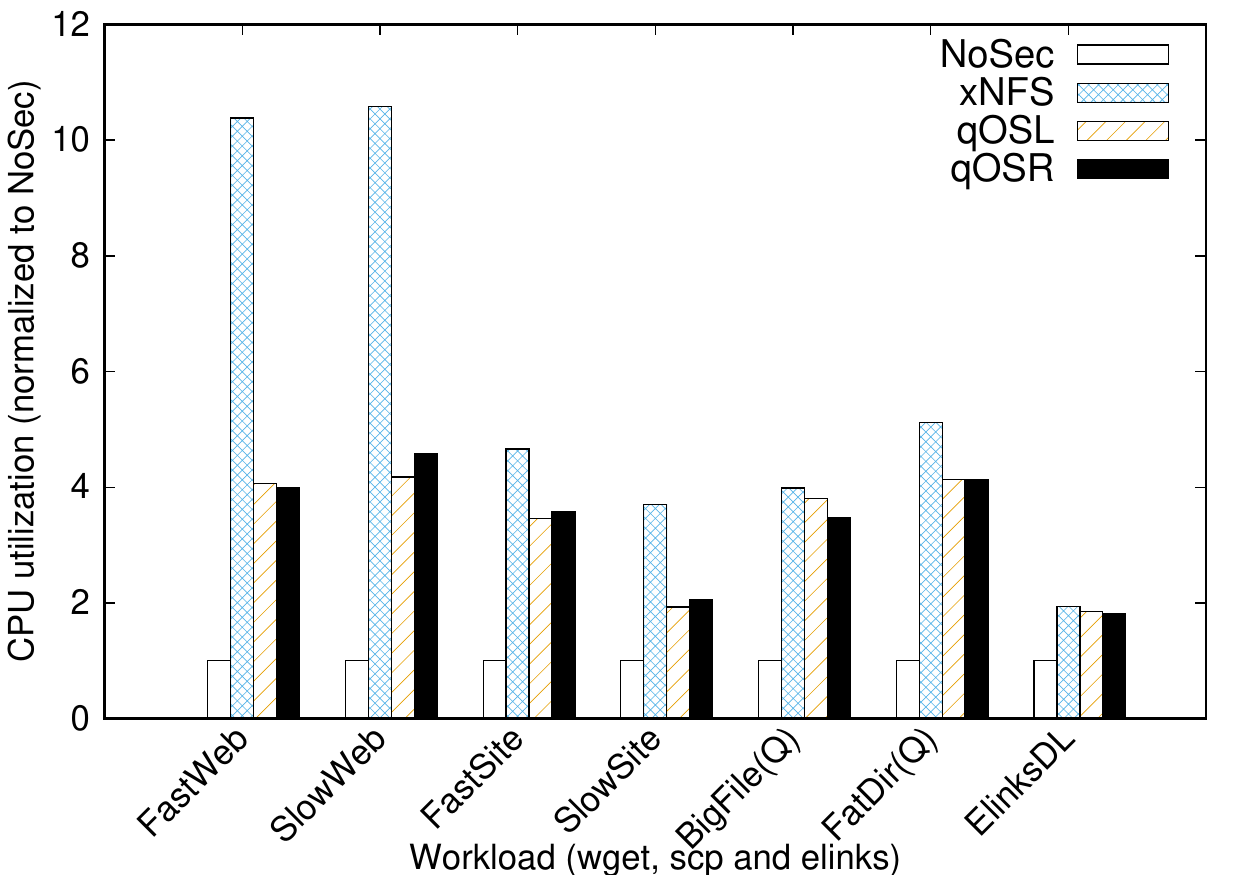}
\caption{Comparative CPU overhead for bulk-data transfer applications using
    different quarantine techniques.}
\label{fig:bulk}
\end{figure}
For batch mode downloads, we download files over network from within the quarantine environment and 
store the files on the host file system (through NFS for $xNFS$ mode or qOS file system APIs).
We used {\tt wget} to download the kernel source from a local hosting site (shown as $FastWeb$ in Figure~\ref{fig:bulk}, 
speed limit = 10MByes/sec), download the same kernel source from the Linux kernel repository (shown as $SlowWeb$,
speed limit = 1MByte/sec), crawl and download a local website (referred to as $FastSite$, speed limit = 10Mbyte/sec)
and crawl a site over internet (referred to as $SlowSite$, speed limit = 1Mbyte/sec).
Further, in this experiment, we have used {\tt scp} in quiet mode to download a single large file of size 
650MB (workload referred to as $BigFile$) and a directory containing a C++ source repository of size 370MB 
(workload referred to as $FatDir$).
Lastly, we used {\tt elinks}, a text-based web browser to download a single file of size $\sim$160MB.

As shown in Figure~\ref{fig:bulk}, VM-based solutions result in a significant CPU overhead (up to 11x) compared 
to the $NoSec$ system because of the associated virtualization overheads.
Note that, as the applications used are not CPU throttling in nature, the download throughput
is the same across all the settings for any given workload.
$qOSL$ and $qOSR$ reduces the CPU overhead by up to 2.5x compared to that of $xNFS$. 
For {\tt wget}, qOS performs significantly better than $xNFS$ leveraging
{\tt write buffering} feature which hides the performance penalty because of 
4096 byte sized writes employed by {\tt wget}. 
For {\tt scp} and {\tt elinks}, in qOSR mode, the CPU overhead is improved by 1.14x and 1.06x compared to the 
$xNFS$ mode. The difference in improvement across {\tt wget} and these two workloads is primarily because
{\tt scp} and {\tt elinks} perform application-level write buffering, which improves the performance of NFS
backed writes.
For the directory transfer workload, qOS modes improve the CPU utilization by a factor of 1.23x
compared to $xNFS$.
This can be because of the NFS overheads associated with a lot of 
meta-data operations for creating files and directories.

\begin{figure}[tp]
\centering
\begin{subfigure}[b]{0.49\columnwidth}
    \includegraphics[width=\textwidth]{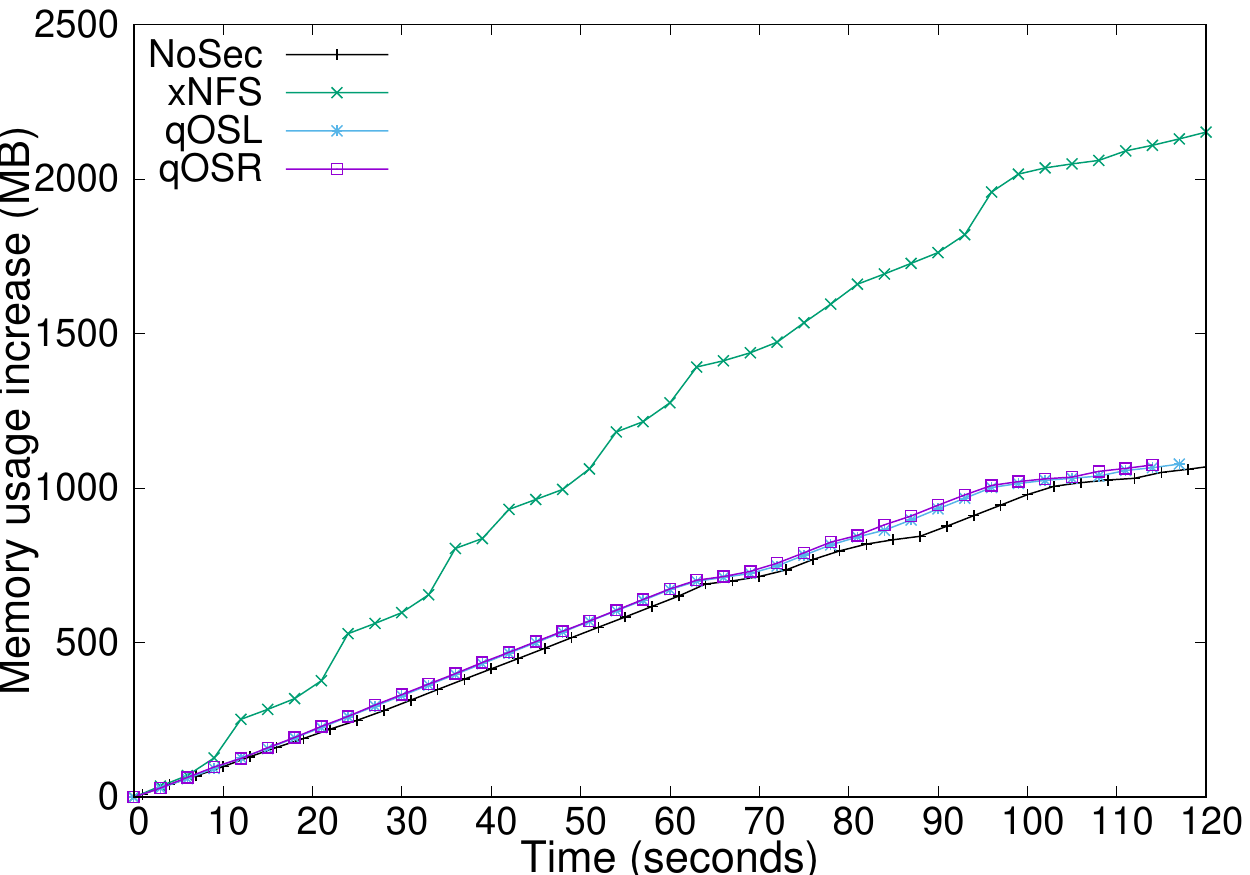}
	\caption{Memory usage ({\tt scp})}
    \label{fig:scp}
  \end{subfigure}
  \hfill
  \begin{subfigure}[b]{0.49\columnwidth}
    \includegraphics[width=\textwidth]{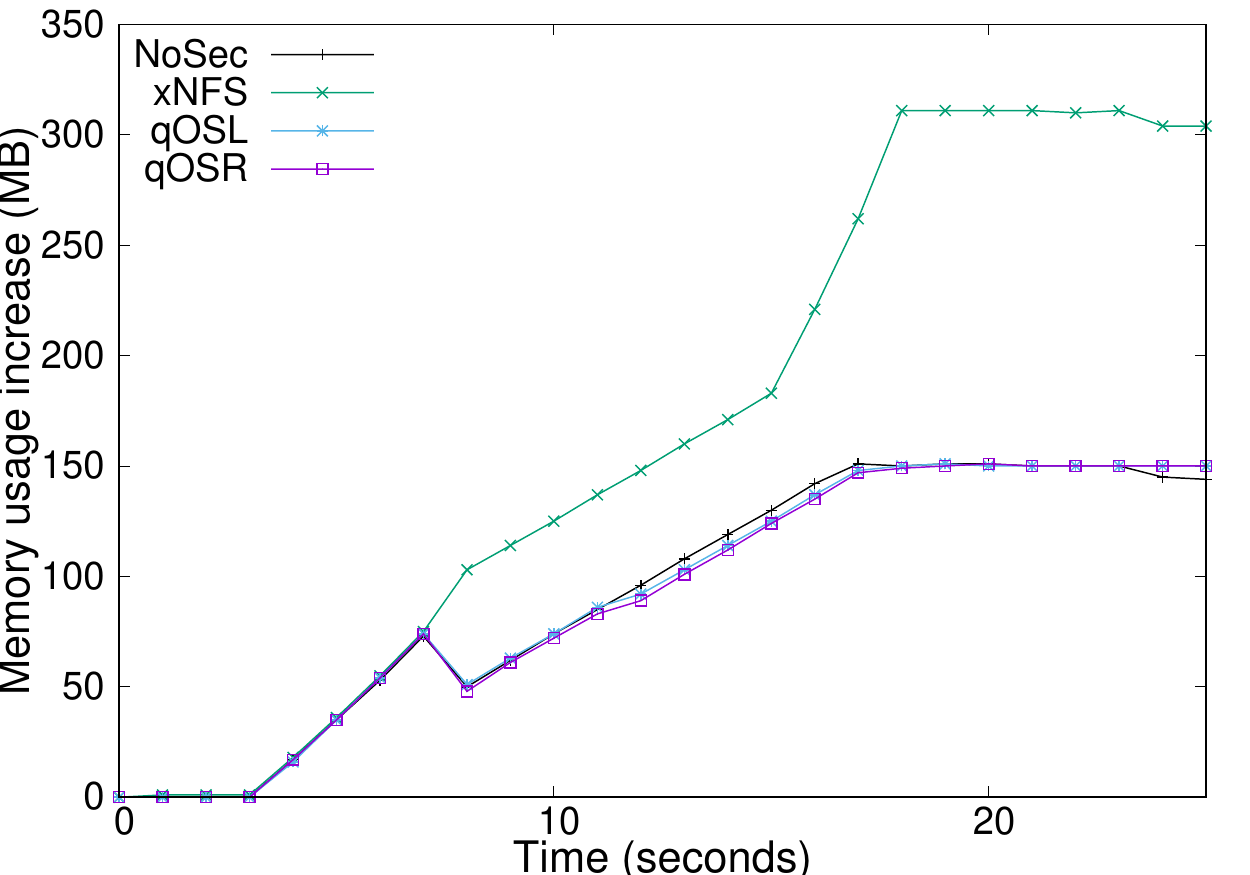}
	  \caption{Memory usage ({\tt elinks})}
    \label{fig:elinks}
  \end{subfigure}
  \caption{Memory usage for the {\tt scp} and {\tt elinks} workloads.
      For qOS modes and xNFS, memory usage is calculated by adding the 
      memory usage increase in the host and the VM.}
\label{fig:memheads}
\end{figure}

To show the system memory utilization during bulk data transfer, we show the incremental memory
allocation during the workload execution for {\tt scp} and {\tt elinks}. 
As shown in Figure~\ref{fig:memheads}, qOSR, and qOSL consume very little additional memory
compared to NoSec mode, while xNFS consumes almost double the amount of memory.
This is primarily because of double caching behavior as the data is cached both in the VM (by the NFS client) 
and the host file system. Note that, in qOS, the disk cache (e.g., page cache, buffer cache, etc.) is maintained only by the host, and there is no penalty for qOS except for a memory-to-memory transfer using the communication channel.
While we have not shown here, {\tt wget} also results in similar memory overheads.
However, as shown earlier, the caching can improve the performance, as shown in the case of GUI applications earlier.

\subsubsection{Interactive applications}

\begin{figure}[tp]
\centering
\begin{subfigure}[b]{0.34\textwidth}
\includegraphics[width=\textwidth]{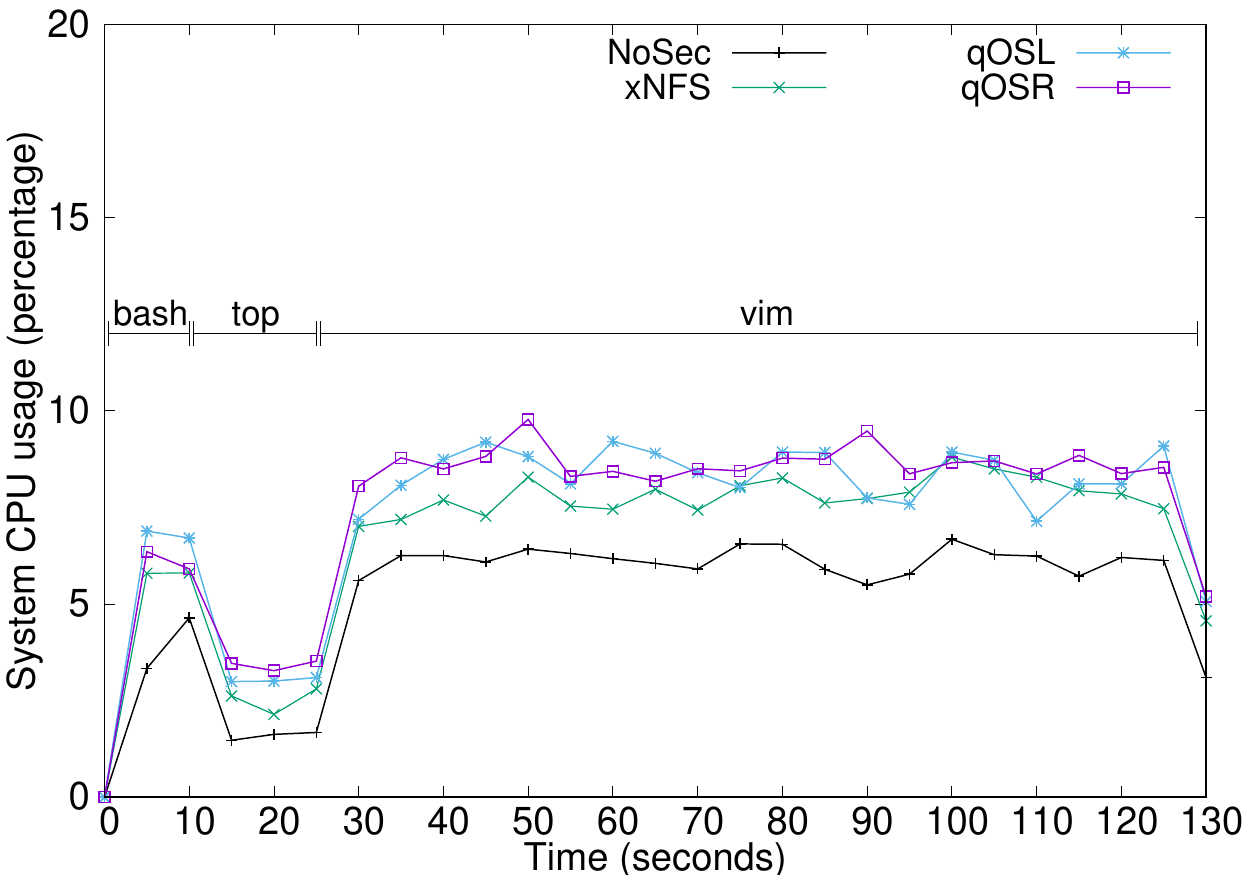}
\caption{Remote shell accessed over SSH.}
\label{fig:intssh}
\end{subfigure}

\begin{subfigure}[b]{0.34\textwidth}
\includegraphics[width=\textwidth]{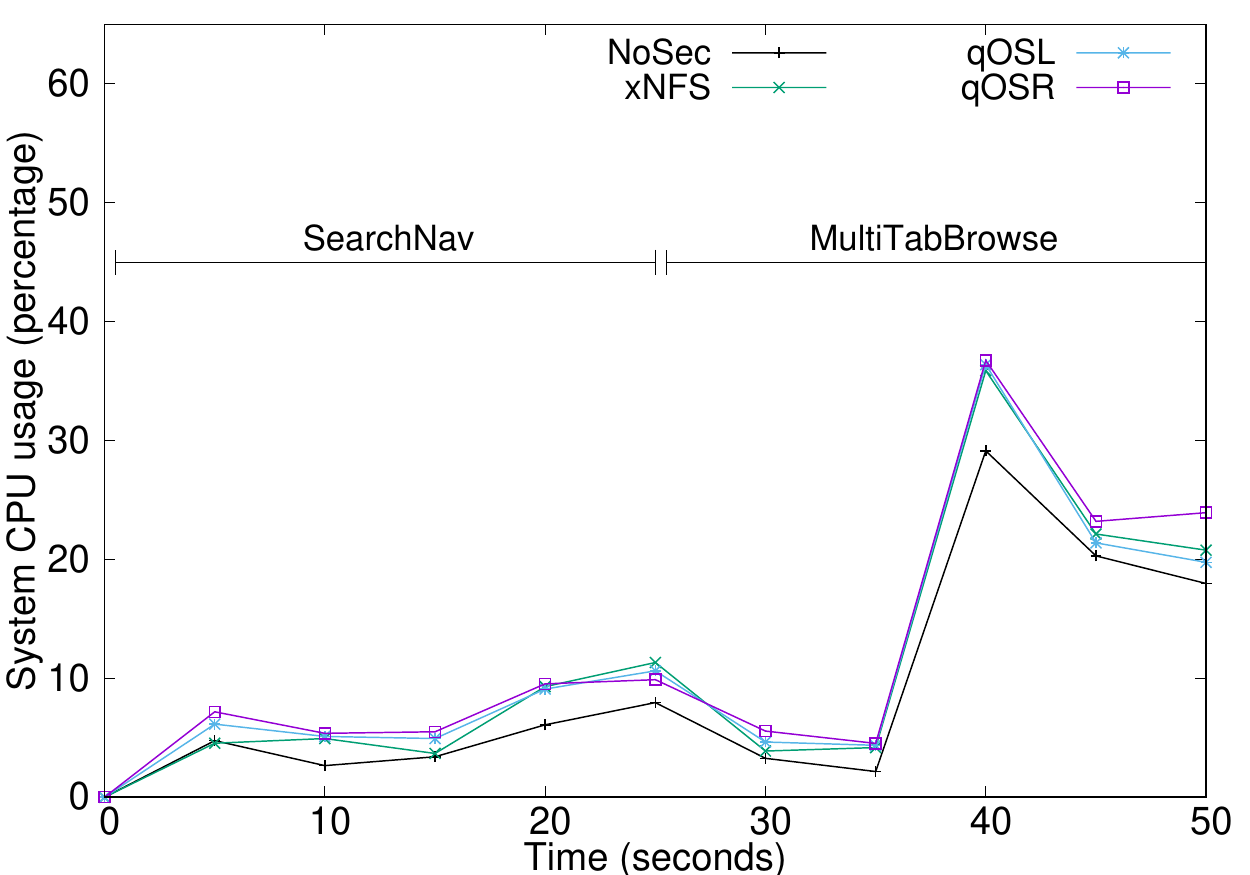}
\caption{Interactive text browser.}
\label{fig:intelinks}
\end{subfigure}
\caption{CPU utilization during the execution of the {\tt scp} and {\tt elinks} workloads with
     different setups.}
\label{fig:cpuheads}
\end{figure}

We have used automated key generation feature~\cite{uinput}
to evaluate the performance of interactive applications like remote shell ({\tt ssh}) and text-based
web browser ({\tt elinks}).
In the Shell workload, we connected to a remote host from the test machine (for $NoSec$, $qOSL$, and $qOSR$) and 
from the VM access through SSH from the test machine (for $xNFS$).
The input trail executes some well known UNIX commands on the remote host, followed by the {\tt top} command.
Finally, it opens a file on the remote host using the VI editor and appends 1000 random characters to it.
With the elinks browser, we search the meaning of a word in google and navigate to the first result.
After that, we open multiple tabs and browse some web sites.
System CPU utilization for both the applications during this experiment is shown in Figure~\ref{fig:cpuheads}.
The qOS modes perform marginally worse (in some cases) compared to the $xNFS$ mode for SSH workload.
The CPU utilization in {\tt elinks} is similar for $xNFS$, $qOSL$ and $qOSR$ which is worse than 
the baseline system by $\sim$10\% in the worst case.

\subsection{Application launch time}

\begin{table}[t]
\small
\centering
\begin{tabular}{c|c|c|c}
\hline
{\bf Application} & {\bf $qOSR$} & {\bf $qOSL$} & {\bf $xNFS$}\\
\hline
\hline
  {$SCP$} & 4.6 & 1.8 & 1\\
  {$Wget$} & 15.8 & 3 & 2.2 \\
  {$Elinks$}  & 24.4 & 5 & 5\\
  {$Evince$}  & 56.6 & 9.6 & 10.4 \\
\hline
\hline
\end{tabular}
\caption{Application start time approx. (in milliseconds).}
\label{tab:startup}
	\vspace{-0.5cm}
\end{table}

To evaluate application launch time performance with different setups, 
we approximate the start-up time of applications as follows.
We start the application with some invalid command-line arguments, which results in application exit, almost immediately, 
giving an estimate of the start-up time for the application. 
The application launch time using the $qOSL$, $qOSR$ and $xNFS$ modes is shown in Table~\ref{tab:startup}. 
We observed that start-up time for $qOSL$ mode is almost the same as $xNFS$ mode, 
as the binary and libraries are loaded from the qVM. 
In $qOSR$ mode, the start-up time increases by a significant factor (by up to 6x) primarily 
due to the overheads of fetching the binaries, library files, etc. from the host system. 
Nevertheless, the absolute values are very small to be noticed by the end-user.

\subsection{Communication Channel Efficacy}

\begin{table}[t]
\small
\centering
\begin{tabular}{c|c|c|c}
\hline
{\bf \#Processes} & {\bf $Average CPU$} & {\bf $Max CPU$} & {\bf \#Messages}\\
\hline
\hline
  {$1$} & 19.56 & 25.0 & 2809204\\
  {$2$} & 26.06 & 29.3 & 5006289 \\
  {$8$} & 29.49 & 42.7 & 5386756 \\
  {$16$} & 47.29 & 57.4 & 6466970\\
  {$32$} & 60.51 & 70.0 & 7565041 \\
  {$64$} & 65.95 & 79.3 & 7990928\\
  {$100$} & 72.38 & 100.0 & 8173783 \\
  {$128$} & 67.09 & 78.6 & 8164545\\
\hline
\hline
\end{tabular}
\caption{Scalability of the split-context on the host.}
\label{tab:eff_expt}
\vspace{-0.5cm}
\end{table}
To find out the responsiveness of the communication channel, we sent 32-byte messages 
from the qOS to the split-context on the host which sends a response without any delay.
We observed an average response time of $\sim$ 24 microseconds.
We found that, interrupt delivery latency and scheduling latency of 
the split-context are the primary contributors.
%

%
%
%
%

To test the scalability of the communication channel, 
we created an application which forks many child processes, 
each of which calls {\tt uname} system call in throttle mode
for one minute. 
For this experiment, we modified the {\tt uname} behavior at the system call 
interceptor to send messages of size 32-bytes to the host through the communication channel 
and wait for their response. 
%
Table~\ref{tab:eff_expt} shows the number of messages processed by 
the HostOS Gateway (executed in the split-context) along with 
average and maximum CPU usage by the split-context during the experiment.
We observe that, with increasing number of child processes in the qVM, 
the split-context can process more messages up to a limit until
maximum CPU utilization is reached (for 100 processes). 

%
%

%
%

%

\section{Discussion}
\label{sec:discussion}

 We reflect upon our experience building qOS, discuss some of the issues which 
 we encountered and possible ways to tackle them as part of the future work. 

\noindent
{\bf Scalable I/O handling:}
The design choice of a fixed 1:M mapping between the host split-context and 
the processes of split-application is primarily to handle denial of service attacks like fork bomb.
However, this can become a bottleneck and result in responsiveness issues. 
%
%
For {\tt select} and {\tt poll} with file handles spanning across guest OS and the host OS, 
any time-slice joggling based solution is required to quickly adapt to strike 
a balance between the CPU overheads and the responsiveness. 
A more attractive alternative can be to adaptively change the number of split-contexts
on the host to address both security and efficiency concerns.

\noindent
{\bf Measuring responsiveness:}
One of the important aspects of GUI applications is responsiveness. 
From our experience of experimenting with different GUI applications, 
we could feel a very good user experience with qOS systems. 
Nevertheless, we can adopt techniques like Deskbench~\cite{deskbench} 
to empirically analyze the responsiveness of qOS, $xNFS$ and $vncNFS$. 

\noindent
{\bf Network access from host:}
  In qOS, the network device is hosted in the qVM.
 %
 Therefore, if the machine does not have multiple interfaces or SRIOV like support, 
 the challenge is to provide connectivity to the host in a secure manner. 
 Techniques like iKernel~\cite{tan2007ikernel} provide a method for creating 
 host network devices through the VM using shared memory channels.
 This can be easily integrated into qOS as the shared channel is an 
 integral part of qOS. 
 %

\noindent
{\bf qOS and VMI:}
The overall security of qOS can be further improved by incorporating
VMI techniques~\cite{rhee2009defeating, hprove}.
In qOS, the shortcomings of black-box VMI can be addressed
by providing useful guest OS statistics and profiling information 
through the qAgent.
We believe this direction has a lot of potential and should be explored in a comprehensive manner.


\section{Related Work}
\label{sec:related_work}

%
Control hijacking is used to exploit vulnerabilities in code and gain control over a victim's machine.
%
%
Attacks like buffer overflow~\cite{morrisworm}, ROP~\cite{rop}, JOP~\cite{jop}, 
CAIN~\cite{cain} etc. exploit control flow integrity vulnerabilities.
Similarly, techniques like data execution prevention (DEP)~\cite{dep}, StackGuard~\cite{cowan1998stackguard}, 
ROPecker~\cite{ropecker} and ASLR~\cite{aslr, lu2015aslr} try to provide better defense against these attacks.
%
%
However, all the defense techniques assume that the administrator to be knowledgeable 
and careful while changing the system configurations.
qOS complements these defenses by addressing the security issues of gullible users.


Another line of defense is to execute potentially harmful applications 
in a restricted environment (a.k.a. sandbox). 
MiniBox ~\cite{minibox} and MBOX~\cite{mbox} are sandbox techniques to protect
the OS and malicious applications from each other. 
MBOX implements a sandbox file system layer above the host file system
to monitor and restrict modifications to the host FS by 
suspicious applications. 
%
Systrace~\cite{provos2003improving} and gVisor~\cite{gvisor,vee2020} use 
system call restriction policies. 
Proxos~\cite{proxos} proposes a system similar to qOS to address the problem 
of securing applications from malicious OSes by executing security-critical applications in a sandbox.
The qOS approach differs as we execute potential entry point applications in the quarantine
and restrict access to files created by the potential entry point applications.
%
%
Several previous works like Drawbridge~\cite{drawbridge}, Graphene~\cite{graphene} and 
Xax~\cite{xax} extend the library OS techniques to provide isolation across
application with transparent display.
However, unlike qOS, they lack the file system transparency in favor of better 
isolation.
Bromium-HP Sure Click~\cite{bromium} is an enterprise solution
which executes chromium browser in a micro-VM, hosts all 
downloaded files in the VM file system and the user accesses them
through the browser.
%
%
While Bromium only cocoons the chromium browser and some file types (like PDF),
qOS is a generic design to quarantine any application and files type.

To monitor and protect the guest OS, researchers have proposed 
several techniques at the hypervisor level.
SecVisor~\cite{secvisor} monitors changes to the MMU and IOMMU 
to enforce user-generated policy and protect the kernel 
from modifications by any malicious DMA writes.
NICKLE~\cite{riley2008guest} protects the kernel memory from malicious code 
using a shadow copy of the guest kernel.
qOS complements the above technqiues.
Virtualization is also used to isolate buggy and vulnerable device drivers. 
LeVassuer et al.~\cite{osdi2004} and iKernel~\cite{tan2007ikernel} propose techniques 
to efficiently isolate device drivers using VMs to contain driver bugs 
and enhance system security. 
Techniques like iKernel can be useful for qOS system (\S\ref{sec:discussion}).

%
Apart from monitoring, virtualization has been used to 
isolate security-sensitive applications and the OS using virtual 
machines (e.g., Terra~\cite{garfinkel2003terra}). 
TrustVisor~\cite{mccune2010trustvisor} and Flicker~\cite{mccune2008flicker} combine
virtualization techniques with hardware support to improve operating system security.
%
Responsiveness of remote desktop mechanisms can be improved by coupling them with 
software like VMGL~\cite{vmgl}. VMGL improves the rendering capabilities of 
X11 and VNC based system by leveraging hardware rendering acceleration. 
qOS can take advantage of such hardware features because of its split-design 
while providing improved security at the same time.

\section{Conclusion}
\label{sec:conclusion} 
In this paper, we proposed an end-to-end implementation and evaluation of qOS, 
a generic and user-transparent solution 
to improve the security of unwary users 
by leveraging virtualization techniques. 
%
%
Compared to straw man solutions like $xNFS$ and $vncNFS$, qOS provides
improved security and efficiency in terms of CPU and memory usage.
Further, qOS is flexible enough to incorporate application-specific optimizations.
We demonstrated some common desktop applications to be working seamlessly with 
qOS providing user experience similar to a native system. 
The empirical evaluation of qOS shows promising results by improving the performance 
in orders of magnitude compared to the existing solutions.
qOS can complement existing security techniques like VMI to attain
higher levels of security for end-user devices.
%
%

\bibliographystyle{ACM-Reference-Format}
\bibliography{qos}

\end{document}